\documentclass[journal,twoside]{IEEEtran}

\usepackage{balance}
\usepackage{cite}
\usepackage{graphicx}
\usepackage{setspace}

\def\e{\begin{equation}}
\def\f{\end{equation}}
\def\l#1{\label{eq:#1}}

\def\vec#1{{\bf #1}}

\begin{document}

\title{Modelling of Wave Propagation in Wire Media Using Spatially Dispersive Finite-Difference Time-Domain Method: Numerical Aspects}

\author{Yan~Zhao,~\IEEEmembership{Student~Member,~IEEE,}
        Pavel~Belov,~\IEEEmembership{Member,~IEEE,}
        and~Yang~Hao,~\IEEEmembership{Member,~IEEE}}

\markboth{IEEE Transactions on Antennas and Propagation}{Y.~Zhao \MakeLowercase{\textit{et al.}}: Modelling of Wave Propagation in Wire Media Using FDTD}

\maketitle

\begin{abstract}
The finite-difference time-domain (FDTD) method is applied for
modelling of wire media as artificial dielectrics. Both
frequency dispersion and spatial dispersion effects in wire media
are taken into account using the auxiliary differential equation
(ADE) method. According to the authors' knowledge, this is the first
time when the spatial dispersion effect is considered in
the FDTD modelling. The stability of developed spatially dispersive
FDTD formulations is analysed through the use of von Neumann method
combined with the Routh-Hurwitz criterion. The results show that the
conventional stability Courant limit is preserved using standard
discretisation scheme for wire media modelling. Flat sub-wavelength lenses formed
by wire media are chosen for validation of proposed spatially
dispersive FDTD formulation. Results of the simulations demonstrate
excellent sub-wavelength imaging capability of the wire medium
slabs. The size of the simulation domain is significantly reduced using the modified perfectly matched layer (MPML) which can be placed in close vicinity of the wire medium. It is demonstrated that the reflections from the MPML-wire medium
interface are less than -70 dB, that lead to dramatic improvement
of convergence compared to conventional simulations.
\end{abstract}

\section{Introduction}
\begin{figure}[t]
    \centering
    \includegraphics[width=6.8cm]{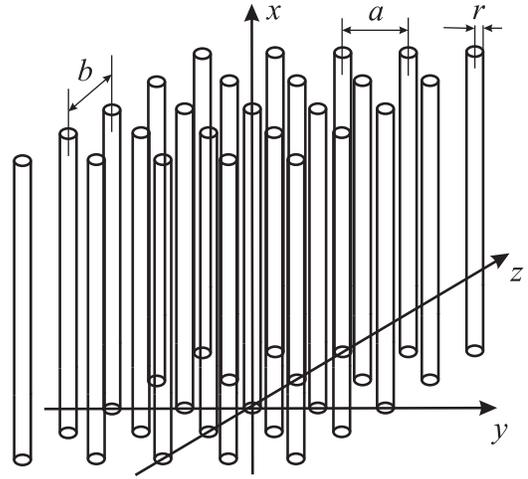}
    \caption{The geometry of the wire medium: a rectangular lattice of parallel ideally conducting thin wires}
    \label{fig_wire_medium}
\end{figure}

The wire medium is an artificial material formed by a
regular lattice of ideally conducting wires (see Fig.
\ref{fig_wire_medium}). The radii of wires are assumed to be small
compared to the lattice periods and the wavelength. The wire medium
has been known for a long time \cite{Brown,Rotman,Pendry} as an
artificial dielectric with plasma-like frequency dependent
permittivity, but only recently it was shown that this dielectric is
non-local and possesses also strong spatial dispersion even at very
low frequencies \cite{Belov1}. Following \cite{Belov1} the wire
medium can be described (if lattice periods are much smaller than the wavelength) as a uniaxial
dielectric with both frequency and spatially dependent effective
permittivity:
\begin{equation}
\overline{\overline{\varepsilon}}=\varepsilon(k,q_{x})\textbf{xx}+\textbf{yy}+\textbf{zz},~\varepsilon(k,q_{x})=\varepsilon_{0}\left(1-\frac{k^{2}_{0}}{k^{2}-q^{2}_{x}}\right),
    \label{eq_permittivity}
\end{equation}
where $k_{0}=\omega_{0}/c$ is the wave number corresponding to the
plasma frequency $\omega_{0}$, $k=\omega/c$ is the wave number of
free space, $c$ is the speed of light, and $q_{x}$ is the component
of wave vector \vec{q} along the wires. The dependence of permittivity
(\ref{eq_permittivity}) on $q_{x}$ represents the spatial dispersion
effect which was not taken into account in the conventional local
uniaxial model of wire medium \cite{Brown,Rotman,Pendry}.

The plasma frequency of wire medium depends on the lattice periods
$a$ and $b$, and on the radius of wires $r$ \cite{BelovJEWA}: \e
k_0^2=\frac{2\pi/(ab)}{\log\frac{\sqrt{ab}}{2\pi r}+F(a/b)}, \l{k0}
\f where \e F(\xi)= -\frac{1}{2}\log \xi+
\sum\limits_{n=1}^{+\infty} \left(\displaystyle
\frac{\mbox{coth}(\pi n \xi)-1}{n}\right)+\frac{\pi}{6}\xi \l{Ffun}.
\f For the commonly used case of the square grid ($a=b$),
$F(1)=0.5275$.

    The finite-difference time-domain (FDTD) method has been widely used for modelling of transient wave propagation in frequency dispersive and non-dispersive media \cite{Taflove}. The existing frequency dispersive FDTD methods can be categorised into three types: the recursive convolution (RC) method \cite{Luebbers1}, the auxiliary differential equation (ADE)
method \cite{Gandhi1} and the Z-transform method \cite{Sullivan1}.
The first frequency dispersive FDTD formulation was developed by
Luebbers \textit{et al.} for the modelling of Debye media
\cite{Luebbers1} using a RC scheme by relating the electric flux
density to the electric field through a convolution integral, and
then discretising the integral as a running sum. The RC approach was
also extended for the study of wave propagation in a Drude material
\cite{Luebbers2}, M-th order dispersive media \cite{Luebbers3}, an
anisotropic magneto-active plasma \cite{Hunsberger}, ferrite
material \cite{Melon}, and the bi-isotropic/chiral media
\cite{Akyurtlu1,Grande,Akyurtlu2}. The ADE method was first used
by Kashiwa and co-workers \cite{Kashiwa1,Kashiwa2} in 1990
for modelling of Debye and Lorentz media. Taflove \cite{Goorjian} soon
extended this model to include effects for nonlinear dispersive
media. Independently, Gandhi \textit{et al.} proposed the ADE method
for treating M-th order dispersive media \cite{Gandhi2,Gandhi3}. The other dispersive formulation based on Z transforms
was proposed by Sullivan \cite{Sullivan1} and then extended to treat nonlinear optical phenomena
\cite{Sullivan2} and chiral media \cite{Demir}. In \cite{Feise}, Feise \textit{et
al.} compared the ADE and Z transform methods and applied the
pseudo-spectral time-domain technique for the modelling of
backward-wave metamaterials to avoid the numerical artifact due to
the staggered grid in FDTD. Lu \textit{et al.} used the effective medium dispersive FDTD method for modelling of the layered left-handed metamaterial \cite{Lu} and demonstrated the zero phase-delay transmission phenomenon. Recently, Lee \textit{et al.} applied the
piecewise linear RC (PLRC) method through an effective medium
approach for modelling of the left-handed materials using the similarity of its effective permittivity and permeability functions to Lorentz material model \cite{Lee}.

    The simulation of wire medium can be performed either by modelling the actual structures i.e. parallel wires, or through the effective medium approach. However, in order to accurately model thin wires in FDTD, either enough fine mesh is required or the conformal FDTD \cite{Hao} needs to be used. In this paper, the effective medium approach is used and the wire medium is modelled as a dielectric material with permittivity of the form (\ref{eq_permittivity}). Due to the similarity of the frequency and spatial dispersion effects in (\ref{eq_permittivity}) with Drude material model, the ADE method can be directly applied for the spatially dispersive FDTD modelling.

\section{Dispersive FDTD Formulations}
    Using the Eq. (\ref{eq_permittivity}) the wire media can be
modelled in FDTD as a frequency and spatially dispersive dielectric.
In order to take into account the dispersive properties of materials
in FDTD modelling, the electric flux density is introduced into the
standard FDTD updating equations. Since the $x$-component of
electric flux density $D_{x}(\omega,q_{x})$ is related to the
$x$-component of the electric field intensity $E_{x}(\omega,q_{x})$
in the spectral (frequency-wave vector) domain as
\begin{equation}
    D_{x}(\omega)=\varepsilon(\omega,q_{x})E_x(\omega),
    \label{eq_D_E}
\end{equation}
one can write that
\begin{equation}
\left(k^{2}-q^{2}_{x}\right)D_{x}+\left(q^{2}_{x}-k^{2}+k^{2}_{0}\right)\varepsilon_{0}E_{x}=0.
    \label{eq_D_E2}
\end{equation}
This equation allows to obtain the constitutive relation in the time-space domain in the following form:
\begin{equation}
\left(\frac{\partial^{2}}{\partial x^{2}}-\frac{1}{c^{2}}\frac{\partial^{2}}{\partial t^{2}}\right)D_{x}+\left(\frac{1}{c^{2}}\frac{\partial^{2}}{\partial t^{2}}-\frac{\partial^{2}}{\partial x^{2}}+k^{2}_{0}\right)\varepsilon_{0}E_{x}=0,
    \label{eq_D_E3}
\end{equation}
using inverse Fourier transformation and the following rules:
\begin{eqnarray}
k^{2}\rightarrow-\frac{1}{c^{2}}\frac{\partial^{2}}{\partial t^{2}},
\qquad q^{2}_{x}\rightarrow-\frac{\partial^{2}}{\partial x^{2}}\nonumber
    \label{eq_k_q}.
\end{eqnarray}

The Eq. (\ref{eq_D_E3}) relates only $x$-components of the electric flux density and field intensity. The permittivity in both $y$- and $z$-directions is the same as in free space since the wires are assumed to be thin.

The FDTD simulation domain is represented by an equally spaced
three-dimensional (3-D) grid with periods $\Delta_{x}$, $\Delta_{y}$
and $\Delta_{z}$ along $x$-, $y$- and $z$-directions, respectively.
The time step is $\Delta_{t}$. For discretisation of
(\ref{eq_D_E3}), we use the central finite difference operators in
time ($\delta^{2}_{t}$) and space ($\delta^{2}_{x}$) as well as the
central average operator with respect to time ($\mu^{2}_{t}$):
\begin{eqnarray}
\frac{\partial^{2}}{\partial t^{2}}\rightarrow\frac{\delta^{2}_{t}}{\Delta^{2}_{t}},\qquad \frac{\partial^{2}}{\partial x^{2}}\rightarrow\frac{\delta^{2}_{x}}{\Delta^{2}_{x}},\qquad k^{2}_{0}\rightarrow k^{2}_{0}\mu^{2}_{t},\nonumber
    \label{eq_operator1}
\end{eqnarray}
where the operators $\delta^{2}_{t}$, $\delta^{2}_{x}$ and $\mu^{2}_{t}$ are defined as in \cite{Hildebrand}:
\begin{eqnarray}
\lefteqn{\!\!\!\!\!\!\!\!\!\!\!\!\!\!\!\!\!\!\!\!\!\!\!\!\delta^{2}_{t}F|^{n}_{m_{x},m_{y},m_{z}}\equiv F|^{n+1}_{m_{x},m_{y},m_{z}}-2F|^{n}_{m_{x},m_{y},m_{z}}}\nonumber\\
&&~~~+F|^{n-1}_{m_{x},m_{y},m_{z}}\nonumber\\
\lefteqn{\!\!\!\!\!\!\!\!\!\!\!\!\!\!\!\!\!\!\!\!\!\!\!\!\delta^{2}_{x}F|^{n}_{m_{x},m_{y},m_{z}}\equiv F|^{n}_{m_{x}+1,m_{y},m_{z}}-2F|^{n}_{m_{x},m_{y},m_{z}}}\nonumber\\
&&~~~+F|^{n}_{m_{x}-1,m_{y},m_{z}}\nonumber\\
\lefteqn{\!\!\!\!\!\!\!\!\!\!\!\!\!\!\!\!\!\!\!\!\!\!\!\!\mu^{2}_{t}F|^{n}_{m_{x},m_{y},m_{z}}\equiv\Bigl(F|^{n+1}_{m_{x},m_{y},m_{z}}+2F|^{n}_{m_{x},m_{y},m_{z}}}\nonumber\\
&&~~~~~+F|^{n-1}_{m_{x},m_{y},m_{z}}\Bigr)/4
   \label{eq_operators}
\end{eqnarray}
Here $F$ represents the field components; $m_{x},m_{y},m_{z}$ are indices corresponding to a certain discretisation point in the FDTD domain, and $n$ is the number of the time steps. The discretised Eq. (\ref{eq_D_E3}) reads
\begin{equation}
\left(\frac{\delta^{2}_{x}}{\Delta^{2}_{x}}-\frac{1}{c^{2}}\frac{\delta^{2}_{t}}{\Delta^{2}_{t}}\right)D_{x}+\left(\frac{1}{c^{2}}\frac{\delta^{2}_{t}}{\Delta^{2}_{t}}-\frac{\delta^{2}_{x}}{\Delta^{2}_{x}}+k^{2}_{0}\mu^{2}_{t}\right)\varepsilon_{0}E_{x}=0.
    \label{eq_D_E4_approx}
\end{equation}
Note that in (\ref{eq_D_E4_approx}), the discretisation of the term $k^{2}_{0}$ of (\ref{eq_D_E3}) is performed using the central average operator $\mu^{2}_{t}$ in order to guarantee the improved stability. The stability of different discretisation schemes is analysed in details in the next section. The Eq. (\ref{eq_D_E4_approx}) can be written as
\begin{eqnarray}
\lefteqn{\!\biggl(\frac{D_{x}|^{n}_{m_{x}+1,m_{y},m_{z}}-2D_{x}|^{n}_{m_{x},m_{y},m_{z}}+D_{x}|^{n}_{m_{x}-1,m_{y},m_{z}}}{\Delta^{2}_{x}}}\nonumber\\
&&\!\!\!\!\!\!\!\!-\frac{1}{c^{2}}\frac{D_{x}|^{n+1}_{m_{x},m_{y},m_{z}}-2D_{x}|^{n}_{m_{x},m_{y},m_{z}}+D_{x}|^{n-1}_{m_{x},m_{y},m_{z}}}{\Delta^{2}_{t}}\biggr)\nonumber\\
&&\!\!\!\!\!\!\!\!\!\!\!\!\!+\varepsilon_{0}\biggl(\frac{1}{c^{2}}\frac{E_{x}|^{n+1}_{m_{x},m_{y},m_{z}}-2E_{x}|^{n}_{m_{x},m_{y},m_{z}}+E_{x}|^{n-1}_{m_{x},m_{y},m_{z}}}{\Delta^{2}_{t}}\nonumber\\
&&\!~-\frac{E_{x}|^{n}_{m_{x}+1,m_{y},m_{z}}-2E_{x}|^{n}_{m_{x},m_{y},m_{z}}+E_{x}|^{n}_{m_{x}-1,m_{y},m_{z}}}{\Delta^{2}_{x}}\nonumber\\
&&\!\!\!\!\!\!\!\!\!\!\!\!\!+k^{2}_{0}\frac{E_{x}|^{n+1}_{m_{x},m_{y},m_{z}}+2E_{x}|^{n}_{m_{x},m_{y},m_{z}}+E_{x}|^{n-1}_{m_{x},m_{y},m_{z}}}{4}\biggr)=0.
    \label{eq_D_E4}
\end{eqnarray}
Therefore, the updating equation for $E_{x}$ in terms of $E_{x}$ and $D_{x}$ at previous time steps is as follows:
\begin{eqnarray}
\lefteqn{\!\!\!\!E_{x}|^{n+1}_{m_{x},m_{y},m_{z}}=\frac{1}{a_{1x}}\biggl[b_{1x}D_{x}|^{n+1}_{m_{x},m_{y},m_{z}}+b_{2x}D_{x}|^{n}_{m_{x}+1,m_{y},m_{z}}}\nonumber\\
&&~~~~~~~~~~~+b_{3x}D_{x}|^{n}_{m_{x},m_{y},m_{z}}+b_{4x}D_{x}|^{n}_{m_{x}-1,m_{y},m_{z}}\nonumber\\
&&~~~~~~~~~~~+b_{5x}D_{x}|^{n-1}_{m_{x},m_{y},m_{z}}-\Bigl(a_{2x}E_{x}|^{n}_{m_{x}+1,m_{y},m_{z}}\nonumber\\
&&~~~~~~~~~~~+a_{3x}E_{x}|^{n}_{m_{x},m_{y},m_{z}}+a_{4x}E_{x}|^{n}_{m_{x}-1,m_{y},m_{z}}\nonumber\\
&&~~~~~~~~~~~+a_{5x}E_{x}|^{n-1}_{m_{x},m_{y},m_{z}}\Bigr)\biggr]
    \label{eq_D_E_difference}
\end{eqnarray}
with the coefficients given by
\begin{eqnarray}
a_{1x}&\!\!\!\!=&\!\!\!\!-\frac{\varepsilon_{0}}{c^{2}\Delta^{2}_{t}}-\frac{\varepsilon_{0}k^{2}_{0}}{4},~~~~~~~~~b_{1x}=-\frac{1}{c^{2}\Delta^{2}_{t}},\nonumber\\
a_{2x}&\!\!\!\!=&\!\!\!\!\frac{\varepsilon_{0}}{\Delta^{2}_{x}},~~~~~~~~~~~~~~~~~~~~~~~b_{2x}=\frac{1}{\Delta^{2}_{x}},\nonumber\\
a_{3x}&\!\!\!\!=&\!\!\!\!\frac{2\varepsilon_{0}}{c^{2}\Delta^{2}_{t}}-\frac{2\varepsilon_{0}}{\Delta^{2}_{x}}-\frac{\varepsilon_{0}k^{2}_{0}}{2},~~~b_{3x}=\frac{2}{c^{2}\Delta^{2}_{t}}-\frac{2}{\Delta^{2}_{x}},\nonumber\\
a_{4x}&\!\!\!\!=&\!\!\!\!\frac{\varepsilon_{0}}{\Delta^{2}_{x}},~~~~~~~~~~~~~~~~~~~~~~~b_{4x}=\frac{1}{\Delta^{2}_{x}},\nonumber\\
a_{5x}&\!\!\!\!=&\!\!\!\!-\frac{\varepsilon_{0}}{c^{2}\Delta^{2}_{t}}-\frac{\varepsilon_{0}k^{2}_{0}}{4},~~~~~~~~~b_{5x}=-\frac{1}{c^{2}\Delta^{2}_{t}}.\nonumber
    \label{eq_D_E_difference_coeff}
\end{eqnarray}

    In the spatially dispersive FDTD modelling of wire medium, the
calculations of \vec{D} from the magnetic field intensity \vec{H},
and \vec{H} from \vec{E} are performed using Yee's standard FDTD
equations \cite{Taflove}, while $E_{x}$ is calculated from $D_{x}$ using
(\ref{eq_D_E_difference}) and $E_{y}=\varepsilon^{-1}_{0}D_{y}$, $E_{z}=\varepsilon^{-1}_{0}D_{z}$. Note that in
(\ref{eq_D_E_difference}), the central difference approximations in
time (for frequency dispersion) for both $D_{x}$ and $E_{x}$ are
used at position $(m_{x},m_{y},m_{z})$, and the central difference approximations in
space (for spatial dispersion) are used at the time step $n$ in
order to update $E_{x}$ at time step $n+1$. Therefore the storage of $D_{x}$ and
$E_{x}$ at two previous time steps are required.

    At the free space-wire medium interfaces along $x$-direction, the updating equation (\ref{eq_D_E_difference}) includes $D_{x}$ and $E_{x}$ at previous time step in both free space and wire medium. Outside the region of wire medium, the updating equation (\ref{eq_D_E_difference}) reduces to the equation relating $D_{x}$ and $E_{x}$ in free space.

\begin{figure}[t]
    \centering
    \includegraphics[width=7.6cm]{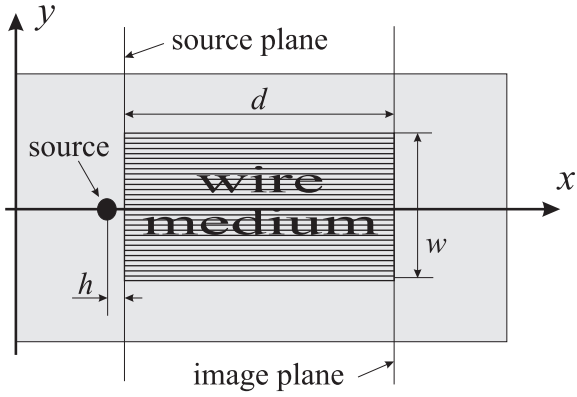}
    \caption{The layout of the computation domain for two-dimensional FDTD simulations}
    \label{fig_simulation_domain}
\end{figure}

        The spatially dispersive FDTD method has been implemented in a two-dimensional case and used for modelling of sub-wavelength imaging provided by a finite sized slab of wire medium \cite{Belov2}. The simulated geometry is illustrated in Fig. \ref{fig_simulation_domain}. Different types of sources are chosen in simulations including a magnetic point source and three equally spaced magnetic point sources in order to demonstrate the sub-wavelength imaging capability of the device. The simulation results are provided in section V, while the stability and numerical dispersion relation for general 3-D case are analysed in the following section.

\section{Stability and Numerical Dispersion Analysis}
\subsection{Stability}
    Previous stability analysis of dispersive FDTD schemes is performed using the von Neumann method and numerical root searching \cite{Petropoulos}. In this paper, the stability of the proposed spatially dispersive FDTD method is analysed using the method combining the von Neumann method with the Routh-Hurwitz criterion as introduced in \cite{Pereda}. The von Neumann method establishes that, for a finite-difference scheme to be stable, all the roots $Z_{i}$ of the stability polynomial $S(Z)$ must be inside of the unit circle in the $Z$-plane (i.e. $\left|Z_{i}\right|\leq1~\forall~i$), where the complex variable $Z$ corresponds to the growth factor of the error and is often called the amplification factor \cite{Pereda}.

    The wire medium is an uniaxial material where the divergence of electric field inside wire medium is non-zero ($\nabla\cdot\vec{E}\neq0$). Therefore, to analyse numerical stability of the proposed spatially dispersive FDTD method, we must start directly with the Maxwell's equations instead of the wave equation as was done in \cite{Pereda} and others for the homogeneous materials.

    Consider the relation between \vec{D} and \vec{E} directly expressed from Faraday's and Ampere's Laws:
\begin{equation}
    \mu_{0}\frac{\partial^{2}\vec{D}}{\partial t^{2}}+\nabla\times\left(\nabla\times\vec{E}\right)=0,
    \label{eq_wave_equation1}
\end{equation}
where $\mu_{0}$ is the permeability of free space. Expansion of the matrix form of (\ref{eq_wave_equation1}) is
\begin{eqnarray}
\lefteqn{\left[\begin{array}{ccc}
	\mu\frac{\partial^{2}}{\partial t^{2}} & 0 & 0\\
	0 & \mu\frac{\partial^{2}}{\partial t^{2}} & 0\\
	0 & 0 & \mu\frac{\partial^{2}}{\partial t^{2}}
	\end{array}\right]\vec{D}}\nonumber\\
	&&+\left[\begin{array}{ccc}
	\frac{\partial^{2}}{\partial x^{2}}-\Delta & \frac{\partial}{\partial x}\frac{\partial}{\partial y} & \frac{\partial}{\partial x}\frac{\partial}{\partial z}\\
	\frac{\partial}{\partial x}\frac{\partial}{\partial y} & \frac{\partial^{2}}{\partial y^{2}}-\Delta & \frac{\partial}{\partial y}\frac{\partial}{\partial z}\\
	\frac{\partial}{\partial x}\frac{\partial}{\partial z} & \frac{\partial}{\partial y}\frac{\partial}{\partial z} & \frac{\partial^{2}}{\partial z^{2}}-\Delta
	\end{array}\right]\vec{E}=0
  \label{eq_wave_equation1_1}
\end{eqnarray}
where
\begin{eqnarray}
\Delta=\frac{\partial^{2}}{\partial x^{2}}+\frac{\partial^{2}}{\partial y^{2}}+\frac{\partial^{2}}{\partial z^{2}}.\nonumber
\end{eqnarray}
Using the central difference operators, Eq.
(\ref{eq_wave_equation1_1}) can be discretized as
\begin{eqnarray}
\mu_{0}\frac{\delta^{2}_{t}}{\Delta^{2}_{t}}\vec{D}+\left(\begin{array}{ccc}
    \frac{\delta^{2}_{x}}{\Delta^{2}_{x}}-\Theta & \frac{\delta_{x}}{\Delta_x}\frac{\delta_{y}}{\Delta_y} & \frac{\delta_{x}}{\Delta_x}\frac{\delta_{z}}{\Delta_z}\\
    \frac{\delta_{x}}{\Delta_{x}}\frac{\delta_{y}}{\Delta_{y}} & \frac{\delta^{2}_{y}}{\Delta^{2}_{y}}-\Theta & \frac{\delta_{y}}{\Delta_{y}}\frac{\delta_{z}}{\Delta_{z}}\\
    \frac{\delta_{x}}{\Delta_x}\frac{\delta_{z}}{\Delta_z} & \frac{\delta_{y}}{\Delta_{y}}\frac{\delta_{z}}{\Delta_{z}} & \frac{\delta^{2}_{z}}{\Delta^{2}_{z}}-\Theta
    \end{array}\right)\vec{E}=0,
  \label{eq_wave_equation2}
\end{eqnarray}
where
\begin{eqnarray}
\Theta=\frac{\delta^{2}_{x}}{\Delta^{2}_{x}}+\frac{\delta^{2}_{y}}{\Delta^{2}_{y}}+\frac{\delta^{2}_{z}}{\Delta^{2}_{z}},\nonumber
\end{eqnarray}
and $\delta_{y}$ and $\delta_{z}$ are defined in the same way as in \cite{Hildebrand}. In addition to the wave equation, the constitutive relation of wire medium (\ref{eq_D_E4_approx}) must also be considered and can be written in the matrix form:
\begin{eqnarray}
\lefteqn{\left[\begin{array}{ccc}
	\frac{\delta^{2}_{x}}{\Delta^{2}_{x}}-\frac{1}{c^{2}}\frac{\delta^{2}_{t}}{\Delta^{2}_{t}} & 0 & 0\\
	0 & -1 & 0\\
	0 & 0 & -1
	\end{array}\right]\vec{D}}\nonumber\\
	&&+\left[\begin{array}{ccc}
\frac{1}{c^{2}}\frac{\delta^{2}_{t}}{\Delta^{2}_{t}}-\frac{\delta^{2}_{x}}{\Delta^{2}_{x}}+k^{2}_{0} & 0 & 0\\
	0 & 1 & 0\\
	0 & 0 & 1
	\end{array}\right]\varepsilon_{0}\vec{E}=0
  \label{eq_wire_medium_D_E}
\end{eqnarray}

    For stability analysis in accordance to \cite{Pereda}, we substitute the following solution into the discrete equations (\ref{eq_wave_equation2}) and (\ref{eq_wire_medium_D_E}),
\begin{equation}
F|^{n}_{m_{x},m_{y},m_{z}}=\tilde{F}Z^{n}e^{j\left(m_{x}\Delta_{x}\tilde{q}_{x}+m_{y}\Delta_{y}\tilde{q}_{y}+m_{z}\Delta_{z}\tilde{q}_{z}\right)},
    \label{eq_solution_Z}
\end{equation}
where $\tilde{F}$ is a complex amplitude, $Z$ is the complex variable which gives the growth of the error in a time interation and $\tilde{\bf{q}}=\left(\tilde{q}_{x},\tilde{q}_{y},\tilde{q}_{z}\right)^{T}$ is the numerical wave vector of the discrete mode. After simple calculations we obtain
\begin{eqnarray}
\lefteqn{\frac{\mu}{\Delta^{2}_{t}}\left(Z-1\right)^{2}\vec{\tilde{D}}}\\
	&&\!\!\!\!\!\!\!\!\!\!\!\!+4Z\left[\begin{array}{ccc}
	\Phi-\frac{\sin^{2}\theta_{x}}{\Delta^{2}_{x}} & \frac{\sin^{2}\theta_{x}}{\Delta_x}\frac{\sin^{2}\theta_{y}}{\Delta_y} & \frac{\sin^{2}\theta_{x}}{\Delta_x}\frac{\sin^{2}\theta_{z}}{\Delta_z}\\
	\frac{\sin^{2}\theta_{x}}{\Delta_x}\frac{\sin^{2}\theta_{y}}{\Delta_y} & \Phi-\frac{\sin^{2}\theta_{y}}{\Delta^{2}_{y}} & \frac{\sin^{2}\theta_{y}}{\Delta_y}\frac{\sin^{2}\theta_{z}}{\Delta_z}\\
	\frac{\sin^{2}\theta_{x}}{\Delta_x}\frac{\sin^{2}\theta_{z}}{\Delta_z} & \frac{\sin^{2}\theta_{y}}{\Delta_y}\frac{\sin^{2}\theta_{z}}{\Delta_z} & \Phi-\frac{\sin^{2}\theta_{z}}{\Delta^{2}_{z}}
	\end{array}\right]\vec{\tilde{E}}=0\nonumber
  \label{eq_wave_equation3}
\end{eqnarray}
where
\begin{eqnarray}
\Phi=\sum_{\alpha=x,y,z}\frac{\sin^{2}\theta_{\alpha}}{\Delta^{2}_{\alpha}},~~~\theta_{\alpha}=\tilde{q}_{\alpha}\Delta_{\alpha}/2,~\alpha=x,y,z,\nonumber
\end{eqnarray}
and
\begin{eqnarray}
\lefteqn{\left[-\frac{1}{c^{2}\Delta^{2}_{t}}Z^{2}+2\left(\frac{1}{c^{2}\Delta^{2}_{t}}-\frac{2\sin^{2}\theta_{x}}{\Delta^{2}_{x}}\right)Z-\frac{1}{c^{2}\Delta^{2}_{t}}\right]\tilde D_{x}}\nonumber\\
&&\!\!\!\!+\biggl[\left(\frac{1}{c^{2}\Delta^{2}_{t}}+\frac{k^{2}_{0}}{4}\right)Z^{2}+\left(-\frac{2}{c^{2}\Delta^{2}_{t}}+\frac{4\sin^{2}\theta_{x}}{\Delta^{2}_{x}}+\frac{k^{2}_{0}}{2}\right)Z\nonumber\\
&&~~~~~~~~~~~~~~~~~~~~~~~~+\left(\frac{1}{c^{2}\Delta^{2}_{t}}+\frac{k^{2}_{0}}{4}\right)\biggr]\varepsilon_{0}\tilde
E_{x}=0\nonumber
    \label{eq_D_E4_Z}
\end{eqnarray}
\begin{equation}
\tilde{D}_{y}-\varepsilon_{0}\tilde{E}_{y}=0,~~~\tilde{D}_{z}-\varepsilon_{0}\tilde{E}_{z}=0.
\label{eq_D_E4_Z2}
\end{equation}
respectively. The determinant of the system of equations (\ref{eq_wave_equation3}) and (\ref{eq_D_E4_Z2}) provides us with the stability polynomial:
\begin{eqnarray}
\lefteqn{S_{\textit{w}}(Z)=\left(\frac{1}{c^{2}\Delta^{2}_{t}}+\frac{k^{2}_{0}}{4}\right)Z^{4}}\nonumber\\
&&\!\!\!\!\!\!\!\!\!\!+4\left(-\frac{1}{c^{2}\Delta^{2}_{t}}+\frac{\sin^{2}\theta_{x}}{\Delta^{2}_{x}}+\Phi\right)Z^{3}\nonumber\\
&&\!\!\!\!\!\!\!\!\!\!+\Biggl[\frac{6}{c^{2}\Delta^{2}_{t}}-\frac{k^{2}_{0}}{2}-\frac{8\sin^{2}\theta_{x}}{\Delta^{2}_{x}}-8\left(1-\frac{2c^{2}\Delta^{2}_{t}\sin^{2}\theta_{x}}{\Delta^{2}_{x}}\right)\Phi\nonumber\\
&&\!\!+\frac{4k^{2}_{0}c^{2}\Delta^{2}_{t}\sin^{2}\theta_{x}}{\Delta^{2}_{x}}\Biggr]Z^{2}\nonumber\\
&&\!\!\!\!\!\!\!\!\!\!+4\left(-\frac{1}{c^{2}\Delta^{2}_{t}}+\frac{\sin^{2}\theta_{x}}{\Delta^{2}_{x}}+\Phi\right)Z+\left(\frac{1}{c^{2}\Delta^{2}_{t}}+\frac{k^{2}_{0}}{4}\right)
	\label{eq_Sw_Z}
\end{eqnarray}
For simplicity, the terms that will lead to the stability condition in free space (i.e. the conventional stability condition) are omitted from this stability polynomial.

In order to avoid numerical root searching \cite{Petropoulos} for obtaining the stability conditions, the above stability polynomial can be transformed into the $r$-plane using the bilinear transformation
\begin{equation}
    Z=\frac{r+1}{r-1}.
    \label{eq_Z_r}
\end{equation}
The stability polynomial in the $r$-plane becomes as follows:
\begin{eqnarray}
\lefteqn{S_{\textit{w}}(r)=\left(\frac{4c^{2}\Delta^{2}_{t}\sin^{2}\theta_{x}}{\Delta^{2}_{x}}\Phi+\frac{k^{2}_{0}c^{2}\Delta^{2}_{t}\sin^{2}\theta_{x}}{\Delta^{2}_{x}}\right)r^{4}}\nonumber\\
&&~~~~~+\Biggl[k^{2}_{0}+\frac{4\sin^{2}\theta_{x}}{\Delta^{2}_{x}}+4\left(1-\frac{2c^{2}\Delta^{2}_{t}\sin^{2}\theta_{x}}{\Delta^{2}_{x}}\right)\Phi\nonumber\\
&&~~~~~~~~~~-\frac{2k^{2}_{0}c^{2}\Delta^{2}_{t}\sin^{2}\theta_{x}}{\Delta^{2}_{x}}\Biggr]r^{2}\nonumber\\
&&~~~~~+\Biggl[\frac{4}{c^{2}\Delta^{2}_{t}}-\frac{4\sin^{2}\theta_{x}}{\Delta^{2}_{x}}-4\left(1-\frac{c^{2}\Delta^{2}_{t}\sin^{2}\theta_{x}}{\Delta^{2}_{x}}\right)\Phi\nonumber\\
&&~~~~~~~~~~+\frac{k^{2}_{0}c^{2}\Delta^{2}_{t}\sin^{2}\theta_{x}}{\Delta^{2}_{x}}\Biggr]
	\label{eq_Sw_r}
\end{eqnarray}

Building up the Routh table for the above polynomial as in
\cite{Pereda}, we obtain the following stability conditions:
\begin{eqnarray}
\lefteqn{\!\!\!\!\!\!\!\!\!\!\!\!\!\frac{4\sin^{2}\theta_{x}}{\Delta^{2}_{x}}\left(1-c^{2}\Delta^{2}_{t}\Phi\right)+4\left(1-\frac{c^{2}\Delta^{2}_{t}\sin^{2}\theta_{x}}{\Delta^{2}_{x}}\right)\Phi}\nonumber\\
&&~~~~~~~~~~~~~~~+k^{2}_{0}\left(1-\frac{2c^{2}\Delta^{2}_{t}\sin^{2}\theta_{x}}{\Delta^{2}_{x}}\right)\geq0
    \label{eq_inequality1}
\end{eqnarray}
\begin{equation}
4\left(\frac{1}{c^{2}\Delta^{2}_{t}}-\frac{\sin^{2}\theta_{x}}{\Delta^{2}_{x}}\right)\left(1-c^{2}\Delta^{2}_{t}\Phi\right)+\frac{k^{2}_{0}c^{2}\Delta^{2}_{t}\sin^{2}\theta_{x}}{\Delta^{2}_{x}}\geq0.
    \label{eq_inequality2}
\end{equation}
In order to fulfil these conditions, it is enough to fulfil the conventional Courant stability condition \cite{Taflove}:
\begin{equation}
\Delta_{t}\leq\frac{1}{c^{2}\Phi}=\frac{1}{c}\left(\sum_{\alpha=x,y,z}\frac{1}{\Delta^{2}_{\alpha}}\right)^{-1/2}
    \label{eq_Courant_condition}
\end{equation}
Therefore the conventional Courant stability condition \cite{Taflove} is preserved for modelling of wire medium and no additional conditions are required.

\begin{figure}[t]
    \centering
\includegraphics[width=8.8cm]{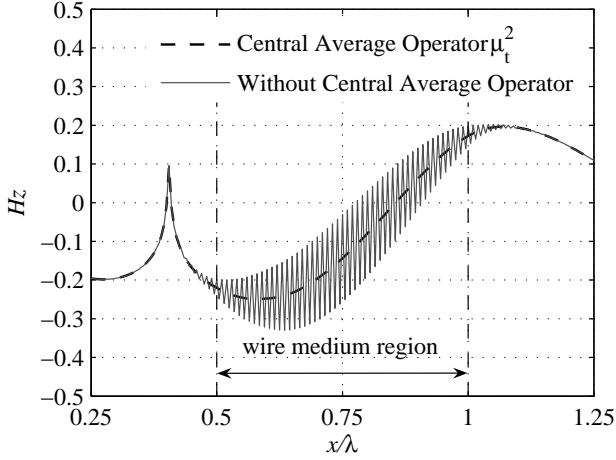}
\caption{Comparison of the internal magnetic field distributions (unit: A/m) in the plane $y=0$ for a wire medium slab excited by a point magnetic source ($d=0.5\lambda$, $w=\lambda$, $h=0.1\lambda$ and $k_{0}/k=4$) calculated using different discretisation schemes: dashed line - the central average operator $\mu^{2}_{t}$, solid line - without using central average operator. The field is plotted at the time step $n=570\Delta_t$, where $\Delta_t$
is chosen as the Courant limit i.e. $\Delta_t=\Delta_x/\sqrt{2}c$, $\Delta_{x}=\lambda/200$.}
    \label{fig_F+S_w=0.5lambda_l=1lambda_cut_stability}
\end{figure}

Note that in the above analysis, the central average operator
$\mu^{2}_{t}$ was used for discretization of the Eq.
(\ref{eq_D_E3}). If we choose the central average operator
$\mu_{2t}$ defined as
\begin{equation}
\mu_{2t}F|^{n}_{m_{x},m_{y},m_{z}}=\left(F|^{n+1}_{m_{x},m_{y},m_{z}}+F|^{n-1}_{m_{x},m_{y},m_{z}}\right)/2,
    \label{eq_mu_2}
\end{equation}
then the stability condition (\ref{eq_inequality1}) remains the same, but (\ref{eq_inequality2}) becomes
\begin{eqnarray}
\lefteqn{4\left(\frac{1}{c^{2}\Delta^{2}_{t}}-\frac{\sin^{2}\theta_{x}}{\Delta^{2}_{x}}\right)\left(1-c^{2}\Delta^{2}_{t}\Phi\right)}\nonumber\\
&&~~~~~+k^{2}_{0}\left(1+\frac{c^{2}\Delta^{2}_{t}\sin^{2}\theta_{x}}{\Delta^{2}_{x}}\right)\geq0
    \label{eq_inequality3}
\end{eqnarray}
which indicates that even less restrictive stability condition
than that for the conventional FDTD method can be reached. However,
if no central average operator is used when discretising the Eq. (\ref{eq_D_E3}), then (\ref{eq_inequality1}) also remains the same, but (\ref{eq_inequality2}) will change to
\begin{equation}
\left(\frac{1}{c^{2}\Delta^{2}_{t}}-\frac{\sin^{2}\theta_{x}}{\Delta^{2}_{x}}\right)\left[4\left(1-c^{2}\Delta^{2}_{t}\Phi\right)-k^{2}_{0}c^{2}\Delta^{2}_{t}\right]\geq0.
    \label{eq_inequality4}
\end{equation}
Therefore, if the plasma frequency used in simulations is too high and the time
step is not properly chosen, then the discretised formulations can
become unstable. Fig.
\ref{fig_F+S_w=0.5lambda_l=1lambda_cut_stability} shows the
comparison of magnetic field distribution using different
discretisation schemes: using the central average operator $\mu^{2}_{t}$
and without using central average operator. It is clearly
shown that after 500 time steps, the instability errors start
appearing from inside the wire medium slab using the latter scheme.

\subsection{Numerical Dispersion}
    The numerical dispersion relation for the wire medium can be found by
evaluating the stability polynomial $S_{w}(Z)$ given by (\ref{eq_Sw_Z}) on the unit circle of the $Z$-plane (i.e. by letting $Z=e^{j\omega\Delta_t}$), and equating the results to zero. After some calculations, the numerical dispersion relation for wire medium is obtained:
\begin{eqnarray}
\lefteqn{\!\!\!\!\!\!\!\!\!\!\!\!\left(\frac{4}{c^{2}\Delta^{2}_{t}}+k^{2}_{0}\right)\sin^{4}\frac{\omega\Delta_{t}}{2}-\left(\frac{4\sin^{2}\theta_{x}}{\Delta^{2}_{x}}+k^{2}_{0}\right)\sin^{2}\frac{\omega\Delta_{t}}{2}}\nonumber\\
&&\!\!\!\!\!\!\!\!\!\!\!\!\!\!\!\!\!\!\!+\frac{k^{2}_{0}c^{2}\Delta^{2}_{t}\sin^{2}\theta_{x}}{\Delta^{2}_{x}}=4\left(\sin^{2}\frac{\omega\Delta_t}{2}-c^{2}\Delta^{2}_{t}\frac{\sin^{2}\theta_{x}}{\Delta^{2}_{x}}\right)\Phi
    \label{eq_dispersion_relation}
\end{eqnarray}
If $\Delta_{\beta}\rightarrow0$, where $\beta=x,y,z,t$, then (\ref{eq_dispersion_relation}) reduces to the continuous dispersion relation for wire medium \cite{Belov1}:
\begin{equation}
\left(q^{2}_{x}-k^{2}\right)\left(q^{2}_{x}+q^{2}_{y}+q^{2}_{z}-k^{2}+k^{2}_{0}\right)=0
  \label{eq_dispersion_relation_continuous}
\end{equation}
The first and second terms of (\ref{eq_dispersion_relation_continuous}) correspond to the transmission line modes (TEM waves with respect to the orientation of wires) and the extraordinary modes (TM waves), see \cite{Belov1} for details. The ordinary modes (TE waves) do not appear in (\ref{eq_dispersion_relation_continuous}) since their contribution was omitted in (\ref{eq_Sw_Z}) for simplicity of calculation.

\section{Perfectly Matched Layer Formulation}
    In 1994, Berenger introduced a nonphysical absorber for terminating the outer boundaries of the FDTD computation domain that has a wave impedance independent of the angle of incidence and frequency. This absorber is called the perfectly matched layer (PML) \cite{Berenger}. The development of PML involves a splitting-field approach and the reflection from a PML boundary is dependent only on the PML's depth and conductivity. In \cite{Zhao}, Berenger's original PML is extended to absorb electromagnetic waves propagating in anisotropic dielectric and magnetic media by introducing the material-independent quantities (electric flux density \vec{D} and magnetic flux density \vec{B}).

PMLs are usually placed at a distance of $\lambda/2$ from any
objects in the simulation domain. In order to reduce the time and
computer memory requirements for simulations as well as to improve
the convergence performance, it is required to place the PML in the
close vicinity of the wire medium. For that purpose, we can follow a
similar approach as in \cite{Zhao} by modifying Berenger's original
PML formulations. In the modified PML for wire medium, \vec{D} is
introduced into the updating equations and the quantities \vec{D}
and \vec{H} are splitted. For example, the updating equation for
$D_{zx}$ becomes
\begin{eqnarray}
\lefteqn{\!\!\!\!\!\!\!\!\!\!D_{zx}|^{n+1}_{m_{x},m_{y},m_{z}}=e^{-\sigma^{D}_{x}\Delta t}D_{zx}|^{n}_{m_{x},m_{y},m_{z}}+\frac{\left(1-e^{-\sigma^{D}_{x}\Delta t}\right)}{-\sigma^{D}_{x}\Delta x}}\nonumber\\
&&~~~\times\biggl[H_{y}|^{n+1/2}_{m_{x}+1/2,m_{y},m_{z}}-H_{y}|^{n+1/2}_{m_{x}-1/2,m_{y},m_{z}}\biggr]
    \label{eq_PML_Dzx}
\end{eqnarray}
and the updating equations for \vec{H} remain the same as in Berenger's original PML \cite{Berenger}. The matching conditions are,
\begin{equation}
    \sigma^{D}_{\alpha}=\sigma^{H}_{\alpha},~\textrm{where}~\alpha=x,y,z,
    \label{eq_matching}
\end{equation}
where $\sigma^{D}_{\alpha}$ and $\sigma^{H}_{\alpha}$ denote electric conductivity and magnetic loss inside the PML, respectively. It should be noted that the difference between the above expressions and Berenger's original PML formulations is that $\varepsilon_{0}$ is not involved in the expression of the theoretical reflection factor $R(\theta)$ for $\sigma^{D}_{\alpha}$ \cite{Zhao}. Furthermore, the updating relation between \vec{D} and \vec{E} in Eq. (\ref{eq_D_E_difference}) must be extended into the PML in order to match wire medium with the modified PML.

\begin{figure}[t]
    \centering
    \includegraphics[width=8.8cm]{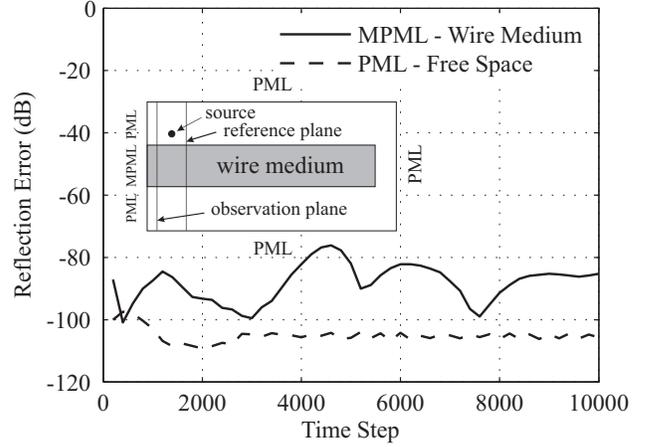}
\caption{Reflection error (in dB) from the MPML-wire medium and
PML-free space interfaces calculated at the observation plane (2
cells away from the PMLs) for a wire medium slab excited by a point
magnetic source ($d=0.5\lambda$, $h=0.1\lambda$ and $k_{0}/k=4$)
plotted as functions of the time step. The wire medium along
$y$-direction is long enough to ensure the wave reflected back from
the far boundary does not reach the reference plane during
calculations.}
    \label{fig_PML_error}
\end{figure}

In order to evaluate the performance of the modified PML for the
wire medium, a 2-D computation domain similar to that in Fig.
\ref{fig_simulation_domain} is chosen except that along
$y$-direction where the wire medium slab is directly terminated by a
ten-cell modified PML (MPML) with a normal theoretical reflection
$R(0)=10^{-5}$ (see sketch in Fig. \ref{fig_PML_error}). From the
other sides the Berenger's original PML is used to truncate the free
space. The source is located close to the edge of the wire medium
slab terminated by MPML and enough far away to the other side of the
slab in order to ensure that the wave reflected back from that side
does not reach the reference plane during calculations. The
observation plane is 2 cells away from the MPML. The reference plane
is located at the same distance from the source as the observation
plane, but from the other side (see sketch in Fig.
\ref{fig_PML_error}). The magnetic fields at the observation and
reference planes are recorded as $H_{\rm{PML}}$ and $H_{\rm{ref}}$,
respectively. The reflection error is defined as
\begin{equation}
    \textrm{Reflection Error (dB)}=20\times\log_{10}\left(\frac{\left|H_{\rm{PML}}-H_{\rm{ref}}\right|}{\left|H_{\rm{max}}\right|}\right),
    \label{eq_reflection_error}
\end{equation}

where $\left|H_{\rm{max}}\right|$ is the maximum value of the
magnetic field at the reference plane. The reflection error is then
calculated for $H_{z}$ and shown in Fig. \ref{fig_PML_error}. For
comparison, the reflection error at the PML-free space interface is
also shown. It is found that the reflections from the PML are less
than -70 dB thus the wire medium is ``perfectly'' matched.

\begin{figure}[t]
    \centering
    \includegraphics[width=8.5cm]{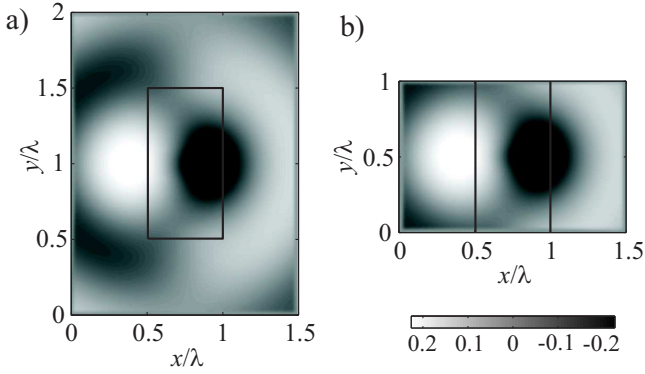}
\caption{(a) Distribution of magnetic field for a finite wire medium
slab excited by a point magnetic source ($d=0.5\lambda$,
$w=\lambda$, $h=0.1\lambda$ and $k_{0}/k=4$). (b) The same slab, but
terminated by a ten-cell MPML at each side along $y$-direction. (unit: A/m)}
    \label{fig_F+S_w=0.5lambda_l=1lambda}
\end{figure}

Fig. \ref{fig_F+S_w=0.5lambda_l=1lambda} shows the comparison of
magnetic field distributions for two cases: using Berenger's
original PML at $\lambda/2$ distance from the slab and using the
modified PML to truncate the wire medium slab directly. In
comparison with Fig. \ref{fig_F+S_w=0.5lambda_l=1lambda}(a), the
simulation domain size for Fig.
\ref{fig_F+S_w=0.5lambda_l=1lambda}(b) is reduced by 50\% and the
convergence is greatly improved since the diffractions from the
corners and edges are avoided in simulations. For the first case,
the reflection error falls below -30 dB after 1000 periods (400,000
time steps), while for the latter one, the convergence is reached
after 100 periods.

\section{Modelling of The Sub-wavelength Imaging}
    In order to validate the proposed spatially dispersive FDTD formulations, we have
chosen flat sub-wavelength lenses formed by wire media
\cite{Belov2}, \cite{Belov3}. Such lenses provide unique opportunity
to transfer images with resolution below classical diffraction
limit. In the present paper we have considered a 2-D case: the
structure is infinite in $z$-direction and electric field is in
$x$-$y$ plane (TM polarization with respect to the orientation of
wires). The following parameters are used in simulations: the
operating frequency is 3.0 GHz (wavelength $\lambda=0.1$ m in free
space) and the plasma frequency of the wire medium is $f_{0}=12.0$
GHz ($k_{0}/k=4$); the FDTD cell size is $\Delta_x=\lambda/200$ with
the time step $\Delta_t=8.33\times10^{-13}$ s according to the
stability criteria (\ref{eq_Courant_condition}); a ten-cell
Berenger's original PML is used to truncate the free space and the
modified PML (MPML) is used to match the wire medium slab; the
thickness of the wire media slab is chosen as $d=\lambda/2$. Three
equally spaced ($\lambda/20$) magnetic point sources with phase
differences of $180^{\circ}$ between neighbouring sources are located at a distance of $\lambda/20$ from the left interface of the wire medium slab.

\begin{figure*}[t]
    \centering
    \includegraphics[width=18cm]{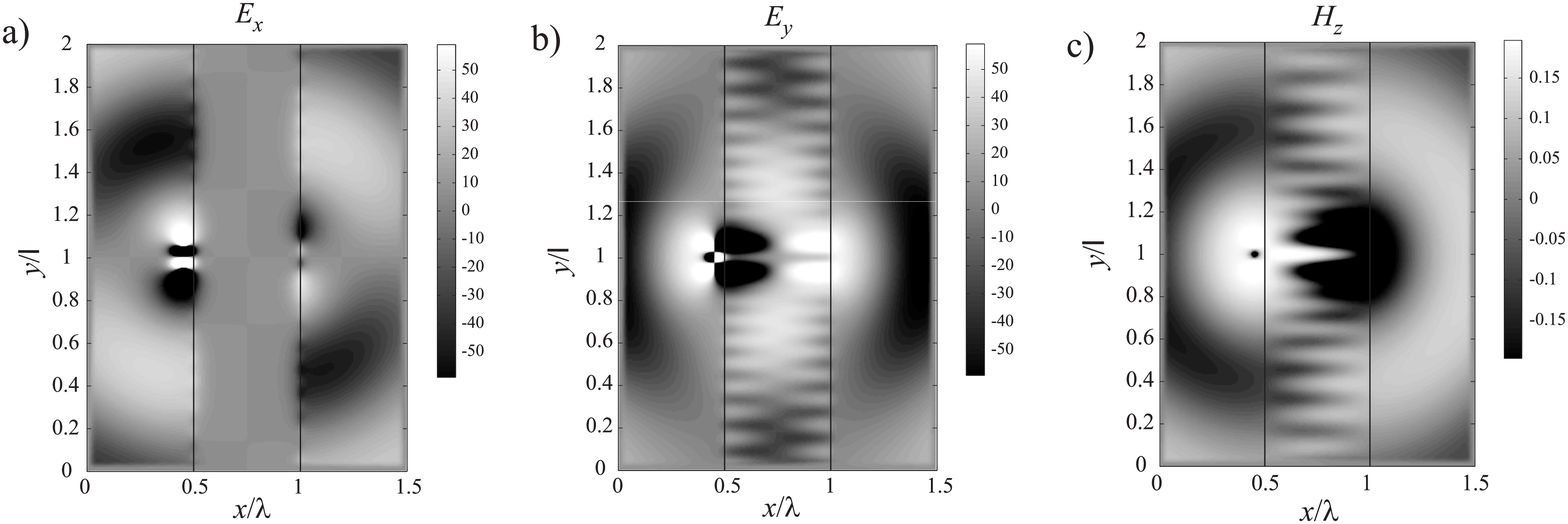}
\caption{The distributions of a) $E_{x}$ (unit: V/m), b) $E_{y}$ (unit: V/m) and c) $H_{z}$ (unit: A/m) for a wire medium slab excited by three equally spaced magnetic
sources with the phase differences equal to $\pi$ ($d=0.5\lambda$,
$h=0.05\lambda$ and $k_{0}/k=4$) directly terminated by a ten-cell
MPML at each side along $y$-direction.}
    \label{ExEyHz}
\end{figure*}

\begin{figure}[t]
    \centering
\includegraphics[width=8cm]{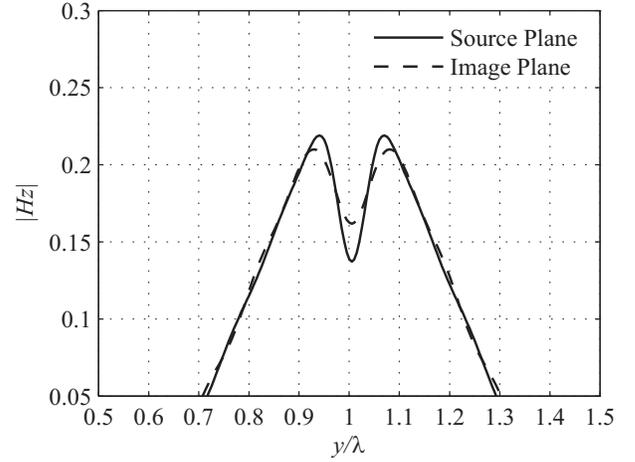}
\caption{Comparison of magnetic field distribution at the source and
image planes for a wire medium slab ($d=0.5\lambda$, $h=0.05\lambda$ and $k_{0}/k=4$) directly terminated by a ten-cell MPML at each side along $y$-direction. (unit: A/m)}
    \label{fig_F+S_w=0.5lambda_l=1lambda_3sources}
\end{figure}

Fig. \ref{ExEyHz} shows the distribution of the $x$- and $y$-
components of electric field as well as $z$- component of magnetic
field in the simulation domain. One can see from Fig. \ref{ExEyHz}(a)
that $x$-component of electrical field penetrates into the wire
medium in the form of extraordinary modes \cite{Belov1} which are
evanescent and decay with distance. That is why this component
vanishes in the center of the slab. The extraordinary modes are
coupled with transmission line modes of wire medium which are
clearly seen in Figs. \ref{ExEyHz}(b) and \ref{ExEyHz}(c). In
accordance to the canalisation principle \cite{Belov2}, the
transmission line modes deliver image from the front interface to
the back one.
Fig. \ref{fig_F+S_w=0.5lambda_l=1lambda_3sources} shows the magnetic
field distribution at the source and image planes (located on the
different sides of wire medium slab as in Fig. \ref{fig_simulation_domain}). It is worth noting that the image is
in phase or out-of-phase with the source if the thickness of wire
medium slab is even or odd integer numbers of $\lambda/2$,
respectively \cite{Belov2}. That is why in the case under
consideration the image appears in out of phase. It can be seen that
in Fig. \ref{fig_F+S_w=0.5lambda_l=1lambda_3sources} the distance
between two maxima is approximately $\lambda/10$ which verifies the
sub-wavelength imaging capability of the wire medium lenses. The
performed simulations confirm that the spatial dispersion in wire
medium is accurately taken into account in the presented spatially
dispersive FDTD model.

\section{Conclusion}
    The spatially dispersive FDTD formulations have been developed for modelling of wave propagation in wire medium as effective dielectrics. The auxiliary differential equation method is used in order to take into account both the spatial and frequency dispersion effects. The stability analysis shows that the conventional Courant stability limit is preserved if the standard central difference approximations and central average operator are used to discretise differential equations. Through the use of modified PML, the wire medium can be ``perfectly'' matched to the absorbing boundaries thus the convergence performance in simulations is greatly improved since the diffractions from corners and edges of the finite sized wire medium slab are avoided. The flat sub-wavelength lenses formed by wire medium are chosen for the validation of developed spatially dispersive FDTD formulations. Numerical simulation results verifies the sub-wavelength imaging capability of wire media. The proposed spatially dispersive FDTD method will demonstrate its distinct value when complex sources are used in the simulation for exploitation of practical applications of wire medium in antenna and microwave engineering.


\begin{thebibliography}{10}
\bibitem{Brown}
J. Brown, \textit{Artificial Dielectrics}. Progress in Dielectrics, vol. 2, pp. 195-225, 1960.

\bibitem{Rotman}
W. Rotman, ``Plasma simulations by artificial dielectrics and parallel-plate media,'' \textit{IRE Trans. Antennas Propag.}, vol. 10, pp. 82-95, 1962.

\bibitem{Pendry}
J. B. Pendry, A. J. Holden, W. J. Steward and I. Youngs, ``Extremely low frequency plasmons in metallic mesostructures,'' \textit{Phys. Rev. Lett.}, vol. 76, pp. 4773-4776, 1996.

\bibitem{Belov1}
P. A. Belov, R. Marques, S. I. Maslovski, I. S. Nefedov, M.
Silveirinha, C. R. Simovski, and S. A. Tretyakov, ``Strong spatial
dispersion in wire media in the very large wavelength limit,''
\textit{Phys. Rev. B}, vol. 67, pp. 113103 (1-4), 2003.

\bibitem{BelovJEWA}
P. A. Belov, S. A. Tretyakov and A. J. Viitanen, ``Dispersion and
reflection properties of artificial media formed by regular lattices
of ideally conducting wires'', \textit{J. Electromagn. Waves Applic.}, vol. 16, pp. 1153-1170, 2002.

\bibitem{Taflove}
A. Taflove, \textit{Computational Electrodynamics: The Finite Difference Time Domain Method}. Norwood, MA: Artech House, 1995.

\bibitem{Luebbers1}
R. Luebbers, F. P. Hunsberger, K. Kunz, R. Standler, and M. Schneider, ``A frequency-dependent finite-difference time-domain formulation for dispersive materials'', \textit{IEEE Trans. Electromagn. Compat.}, vol. 32, pp. 222-227, Aug. 1990.

\bibitem{Gandhi1}
O. P. Gandhi, B.-Q. Gao, and J.-Y. Chen, ``A frequency-dependent finite-difference time-domain formulation for general dispersive media,'' \textit{IEEE Trans. Microwave Theory Tech.}, vol. 41, pp. 658-664, Apr. 1993.

\bibitem{Sullivan1}
D. M. Sullivan, ``Frequency-dependent FDTD methods using Z transforms,'' \textit{IEEE Trans. Antennas Propagat.}, vol. 40, pp. 1223-1230, Oct. 1992.

\bibitem{Luebbers2}
R. J. Luebbers, F. Hunsberger, and K. S. Kunz, ``A frequency-dependent finite-difference time-domain formulation for transient propagation in plasma,'' \textit{IEEE Trans. Antennas Propagat.}, vol. 39, no. 1, pp. 29-34, 1991.

\bibitem{Luebbers3}
R. J. Luebbers and F. Hunsberger, ``FDTD for Nth-order dispersive media,'' \textit{IEEE Trans. Antennas Propagat.}, vol. 40, no. 11, pp. 1297-1301, 1992.

\bibitem{Hunsberger}
F. Hunsberger, R. J. Luebbers, and K. S. Kunz, ``Finite-difference time-domain analysis of gyrotropic media. I: Magnetized plasma,'' \textit{IEEE Trans. Antennas Propagat.}, vol. 40, no. 12, pp. 1489-1495, 1992.

\bibitem{Melon}
C. Melon, P. Leveque, T. Monediere, A. Reineix, and F. Jecko, ``Frequency dependent finite-difference-time-domain formulation applied to ferrite material,'' \textit{Microwave Opt. Technol. Lett.}, vol. 7, no. 12, pp. 577-579, 1994.

\bibitem{Akyurtlu1}
A. Akyurtlu and D. H. Werner, ``BI-FDTD: a novel finite-difference time-domain formulation for modeling wave propagation in bi-isotropic media,'' \textit{IEEE Trans. Antennas Propagat.}, vol. 52, no. 2, pp. 416-425, 2004.

\bibitem{Grande}
A. Grande, I. Barba, A. Cabeceira, J. Represa, P. So, and W. Hoefer, ``FDTD modeling of transient microwave signals in dispersive and lossy bi-isotropic media,'' \textit{IEEE Trans. Microwave Theory Tech.}, vol. 52, no. 3, pp. 773-784, 2004.

\bibitem{Akyurtlu2}
A. Akyurtlu and D. H. Werner, ``A novel dispersive FDTD formulation for modelling transient propagation in chiral metamaterials,'' \textit{IEEE Trans. Antennas Propagat.}, vol. 52, no. 9, pp. 2267-2276, 2004.

\bibitem{Kashiwa1}
T. Kashiwa, N. Yoshida, and I. Fukai, ``A treatment by the finite-difference time-domain method of the dispersive characteristics associated with orientation polarization,'' \textit{Trans. IEICE}, vol. E73, no. 8, pp. 1326-1328, 1990.

\bibitem{Kashiwa2}
T. Kashiwa and I. Fukai, ``A treatment by the FD-TD method of the dispersive characteristics associated with electronic polarization,'' \textit{Microwave Opt. Technol. Lett.}, vol. 3, no. 6, pp. 203-205, 1990.

\bibitem{Goorjian}
P. M. Goorjian and A. Taflove, ``Direct time integration of Maxwell's equations in nonlinear dispersive media for propagation and scattering of femtosecond electromagnetic solitons,'' \textit{Optics Lett.}, vol. 17, no. 3, pp. 180-182, 1992.

\bibitem{Gandhi2}
O. P. Gandhi, B. Q. Gao, and J. Y. Chen, ``A frequency-dependent finite-difference time-domain formulation for induced current calculations in human beings,'' \textit{Bioelectromagnetics}, vol. 13, no. 6, pp. 543-556, 1992.

\bibitem{Gandhi3}
O. P. Gandhi, B. Q. Gao, and J. Y. Chen, ``A frequency-dependent finite-difference time-domain formulation for general dispersive media,'' \textit{IEEE Trans. Microwave Theory Tech.}, vol. 41, no. 4, pp. 658-665, 1993.

\bibitem{Sullivan2}
D. M. Sullivan, ``Nonlinear FDTD formulations using Z transforms,'' \textit{IEEE Trans. Microwave Theory Tech.}, vol. 43, no. 3, pp. 676-682, 1995.

\bibitem{Demir}
V. Demir, A. Z. Elsherbeni, and E. Arvas, ``FDTD formulation for dispersive chiral media using the Z transform method,'' \textit{IEEE Trans. Antennas Propagat.}, vol. 53, no. 10, pp. 3374-3384, 2005.

\bibitem{Feise}
M. W. Feise, J. B. Schneider and P. J. Bevelacqua, ``Finite-difference and pseudospectral time-domain methods applied to backward-wave metamaterials,'' \textit{IEEE Trans. Antennas Propagat.}, vol. 52, no. 11, pp. 2955-2962, 2004.

\bibitem{Lu}
L. Lu, Y. Hao, and C. Parini, ``Dispersive FDTD characterisation of no phase-delay radio transmission over layered left-handed meta-materials structure,'' \textit{IEE Proceedings-Science Measurement and Technology}, 151 (6): 403-406 Nov 2004.

\bibitem{Lee}
J.-Y. Lee, J.-H. Lee. H.-S. Kim, N.-W. Kang and H.-K. Jung, ``Effective medium approach of left-handed material using a dispersive FDTD method,'' \textit{IEEE Trans. Magnetics}, vol. 41, no. 5, pp. 1484-1487, 2005.

\bibitem{Hao}
Y. Hao and C. J. Railton, ``Analyzing Electromagnetic Structures with Curved Boundaries on Cartesian FDTD Meshes,'' \textit{IEEE Transactions on Microwave Theory \& Techniques}, vol.46, no.1, pp.82-8, Jan. 1998.

\bibitem{Hildebrand}
F. B. Hildebrand, \textit{Introduction to Numerical Analysis}. New York: Mc-Graw-Hill, 1956.

\bibitem{Belov2}
P. A. Belov, C. R. Simovski and P. Ikonen, ``Canalisation of
sub-wavelength images by electromagnetic crystals,'' \textit{Phys.
Rev. B}, vol. 71, pp. 193105 (1-4), 2005.

\bibitem{Petropoulos}
P. G. Petropoulos, ``Stability and phase error analysis of FD-TD in dispersive dielectrics,'' \textit{IEEE Trans. Antennas Propagat.}, vol. 42, pp. 62-69, Jan. 1994.

\bibitem{Pereda}
A. Pereda, L. A. Vielva, A. Vegas and A. Prieto, ``Analyzing the stability of the FDTD technique by combining the von Neumann method with the Routh-Hurwitz criterion,'' \textit{IEEE Trans. Microwave Theory Tech.}, vol. 49, no. 2, pp. 377-381, Feb. 2001.

\balance

\bibitem{Berenger}
J. R. Berenger, ``A perfectly matched layer for the absorption of electromagnetic waves,'' \textit{J. Computat. Phys.}, vol. 114, pp. 185-200, Oct. 1994.

\bibitem{Zhao}
A. P. Zhao, ``Generalized-material-independent PML absorbers used for the FDTD simulation of electromagnetic waves in 3-D arbitrary anisotropic dielectric and magnetic media,'' \textit{IEEE Trans. Microwave Theory Tech.}, vol. 46, no. 10, part 1, pp. 1511-1513, Oct. 1998.

\bibitem{Belov3}
P. A. Belov, Y. Hao and S. Sudhakaran, ``Sub-wavelength imaging by wire media,'' \textit{Phys. Rev. B}, vol. 73, pp. 033108 (1-4), 2006.


\end{thebibliography}
\end{document}